\documentclass[12pt,fleqn]{article}
\usepackage{color,amsmath,epsfig}

\newcommand{\be}{\begin{eqnarray}}
\newcommand{\ee}{\end{eqnarray}}
\newcommand{\nee}{\nonumber\end{eqnarray}}
\newcommand{\nn}{\nonumber}

\newcommand{\mbf}      {\boldmath}

\newcommand{\smaf}[2]  {{\textstyle \frac{#1}{#2} }}

\newcommand{\eq}[1]  {\mbox{(\ref{eq:#1})}}
\newcommand{\fig}[1] {\mbox{Fig.~\ref{fig:#1}}}
\newcommand{\Fig}[1] {\mbox{Figure~\ref{fig:#1}}}

\def\b               {\beta}
\def\d               {\delta}

\def\G               {\Gamma}

\def\x               {\chi}

\def\ti              {\tilde}

\def\st              {\ti t}
\def\sb              {\ti b}
\def\ch              {\ti \x^\pm}

\def\nt              {\ti \x^0}
\def\sg              {\ti g}

\newcommand{\mst}[1]   {m_{\st_{#1}}}
\newcommand{\msb}[1]   {m_{\sb_{#1}}}

\newcommand{\mhp}      {m_{H^+}}
\newcommand{\msg}      {m_{\ti g}}

\newcommand{\gsim}{\;\raisebox{-0.9ex}
           {$\textstyle\stackrel{\textstyle >}{\sim}$}\;}
\newcommand{\lsim}{\;\raisebox{-0.9ex}{$\textstyle\stackrel{\textstyle<}
           {\sim}$}\;}

\renewcommand{\Re}{{\rm Re}}

\begin{document}

%==============================================================================
%  Title
%==============================================================================

\begin{center}

{\large\bf\boldmath
    CP violation in charged Higgs boson decays\\ 
    in the MSSM with complex parameters}\\[5mm]

E. Christova, H. Eberl, W. Majerotto, S. Kraml\,\footnote{Speaker}\\[5mm]

{\it Contribution to SUSY02, 
10th International Conference on Supersymmetry 
and Unification of Fundamental Interactions, 
17--23 June 2002, DESY Hamburg, Germany.}

\end{center}

\begin{abstract}
Supersymmetric loop contributions can lead to different
decay rates of $H^+\to t\bar b$ and $H^-\to b\bar t$.
We calculate the asymmetry
$\d^{CP} = [\G(H^+\to t\bar b)-\G(H^-\to b\bar t)]\,/\,
           [\G(H^+\to t\bar b)+\G(H^-\to b\bar t)]$
at next-to-leading order in the MSSM with complex parameters.
We analyse the parameter dependence
of $\d^{CP}$ with emphasis on the phases of $A_t$ and $A_b$.
It turns out that the most important contribution comes from the
loop with stop, sbottom, and gluino. If this contribution is present,
$\d^{CP}$ can go up to 10\,--\,15\%.
\end{abstract}

%==============================================================================
%\section{Introduction}
%==============================================================================

Several talks at this conference dicussed the issue of CP-violating
phases in the supersymmetric (SUSY) Lagrangian. In the Minimal Supersymmetric 
Standard Model (MSSM), the higgsino parameter $\mu$ in the superpotential,
two of the soft SUSY-breaking Majorana gaugino masses $M_i$ ($i=1,2,3$),
and the trilinear couplings $A_f$ (corresponding to a fermion $f$)
can have physical phases, which cannot be rotated away
without introducing phases in other couplings~\cite{Dugan:1984qf}.
From the point of view of baryogenesis, one might hope that these
phases are large~\cite{Carena:1997ki}.
On the other hand, the experimental limits on electron and neutron 
electric dipole moments (EDMs)~\cite{Altarev:cf},
$|d^e| \le 2.15\times 10^{-13}$~e/GeV,
$|d^n| \le 5.5\times 10^{-12}$~e/GeV, place severe constraints
on the phase of $\mu$, $\phi_\mu < {\cal O}(10^{-2})$~\cite{Nath:dn},
for a typical SUSY mass scale of the order of a few hundred GeV.
A larger $\phi_\mu$ imposes fine-tuned relationships between this
phase and other SUSY parameters~\cite{Ibrahim:1997nc}.
Phases of the trilinear couplings of the third generation $A_{t,b,\tau}$ 
are much less constrained and can lead to significant CP-violation
effects, especially in top quark physics~\cite{Atwood:2000tu}. 
Moreover, they can have a significant influence on the phenomenology 
of stop, sbottoms, and staus \cite{Bartl:2002hi}. 
Phases of $\mu$ and $A_{t,b,\tau}$ also affect the Higgs sector
in a relevant way. Although the Higgs potential of the MSSM is
invariant under CP at tree level, at loop level CP is sizeably
violated by complex couplings~\cite{Pilaftsis:1998pe,CEPW00,Carena:2001fw}.
As a consequence, the three neutral mass eigenstates $H^0_l$ ($l=1,2,3$) 
are superpositions of the CP eigenstates $h^0$, $H^0$, and $A^0$.

In this contribution, we discuss CP violation in the decays of charged 
Higgs bosons within the MSSM with complex parameters. 
In particular, we concentrate on the decays into top and 
bottom quarks. Here SUSY loop contributions can lead to a CP-violating 
asymmetry
\begin{equation}
  \d^{CP} = \frac{\G\,(H^+\to t\bar b)-\G\,(H^-\to b\bar t)}
                 {\G\,(H^+\to t\bar b)+\G\,(H^-\to b\bar t)}
\label{eq:defDCP}
\end{equation}
which could be measured in a counting experiment.
We calculate $\d^{CP}$ at the one-loop level in the MSSM with phases 
and discuss its parameter dependence. 
Analogous asymmetries can of course be obtained for other decay 
channels of $H^\pm$, such as $H^\pm\to \tau\nu$, $H^\pm\to\ch\nt$, 
$H^\pm\to\st\sb$.

%==============================================================================
%\section{The \boldmath $H^\pm$ decay}
%==============================================================================

We first discuss the basic formulae for the $H^\pm\to tb$ decays.
The decay widths at tree level are given by
\begin{equation}
  \G^{\,0}\,(H^\pm \to tb) = \frac{3\kappa}{16\pi\mhp^3}
  \left[ (\mhp^2 - m_t^2 - m_b^2)(y_t^2+y_b^2)
         - 4 m_t m_b y_t y_b \right] \,,
\end{equation}
where $\kappa = \kappa(\mhp^2,m_t^2,m_b^2)$, 
$\kappa(x,y,z)=[(x-y-z)^2-4yz]^{1/2}$ and 
\begin{equation}
  y_t = h_t\cos\b \,, \qquad
  y_b = h_b\sin\b \,,
\end{equation}
with $h_t$ and $h_b$ the top and bottom Yukawa couplings. 
Since there is no CP violation at tree level, 
$\G^{\,0}(H^+ \to t\bar b) = \G^{\,0}(H^- \to b\bar t)$. 
At one loop, however, we have 
\begin{equation}
  y_i\to Y_i^{\pm} = y_i^{} + \d Y_i^{\pm} \quad (i=t,b) \,,
  \label{eq:Yi}
\end{equation}
and thus 
\begin{align}
  \G\,(H^\pm \to tb) & =  \frac{3\kappa}{16\pi\mhp^3}
  \left[ (\mhp^2 - m_t^2 - m_b^2)(y_t^2+y_b^2
    + 2y_t\,\Re\,\d Y_t^\pm + 2y_b\,\Re\,\d Y_b^\pm )\right.\nn\\
  & \hspace{24mm}\left. -\, 4 m_t m_b (y_t y_b
    + y_t\,\Re\,\d Y_b^\pm + y_b\,\Re\,\d Y_t^\pm ) \right] \,, 
\label{eq:NLO}
\end{align}
where $\d Y_i^+$ ($i=t,b$) stands for the decay of $H^+$ and 
$\d Y_i^-$ for the decay of $H^-$. 
These form factors have, in general, 
both CP-invariant and CP-violating contributions:
\begin{equation}
  \d Y_i^\pm = \d Y_i^{inv} \pm \smaf{1}{2}\,\d Y_i^{CP}\,.
  \label{eq:dYi}
\end{equation}
Both the CP-invariant and the CP-violating contributions have real and
imaginary parts.
CP invariance implies $\Re\,\d Y_i^+=\Re\,\d Y_i^-$.
Using eqs.~\eq{NLO} and \eq{dYi}, we can write the CP-violating asymmetry 
$\d^{CP}$ of eq.~\eq{defDCP} as %\\
%\vspace*{-4mm}
\begin{equation}
  \d^{CP} =
  \frac{\Delta\,(y_t\,\Re\,\d Y_t^{CP} + y_b\,\Re\,\d Y_b^{CP})
        - 2 m_t m_b (y_t\,\Re\,\d Y_b^{CP} + y_b\,\Re\,\d Y_t^{CP})}
       {\Delta\,(y_t^2+y_b^2) - 4 m_t m_b\,y_t^{} y_b^{} }\,,
\label{eq:DCP}
\end{equation}
where $\Delta = \mhp^2 - m_t^2 - m_b^2 $. 
$\d^{CP}$ gets contributions from loop exchanges of $\st$, $\sb$, $\sg$, 
$\ch$, $\nt$, $W$, and neutral Higgs bosons. 
In principle, there would also be a contribution due to $\ti\nu$ 
and $\ti\tau$ exchange, which can, however, be neglected in our study. 
The relevant Feynman diagrams 
are shown in Fig.~1. The explicit expressions for $\d Y_{t,b}^{CP}$ due 
to these diagrams are given in \cite{Christova:2002ke}.
Of course, the various diagrams contribute to $\d^{CP}$ only 
if they have absorptive parts. Here note that the diagram with $WH^0b$ 
always contributes, since $m_t > m_W+m_b$. 
The dominant contribution, however, comes from the $\st\sb\sg$ loop, 
provided the channel $H^\pm\to \st\sb$ is open.

\begin{figure}[ht!]
\begin{center} {\setlength{\unitlength}{1mm}
\begin{picture}(160,57)
%\put(0,0){\framebox(160,57){}}
\put(-12,-165){\mbox{\epsfig{figure=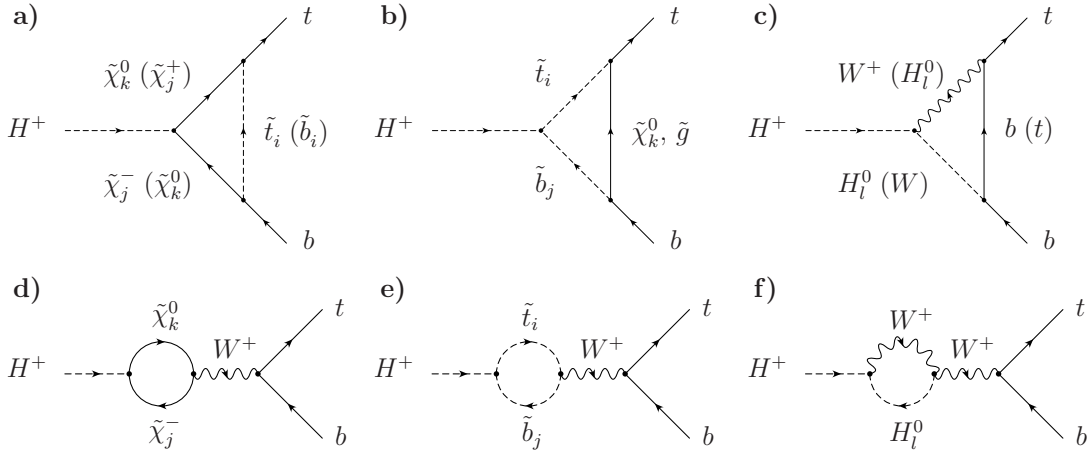,width=18cm}}}
\end{picture} }
\end{center}
\caption{Sources for CP violation in $H^+\to t\bar b$ decays
at 1-loop level in the MSSM with complex couplings
($i,j=1,2;$ $k=1,...,4$; $l=1,2,3$).
\label{fig:feyngraphs}}
\end{figure}

%==============================================================================
%\section{Numerical results}
%==============================================================================

Let us now turn to the numerical analysis. 
In order not to vary too many parameters, we fix part
of the parameter space at the electroweak scale by the choice
\begin{align}
   & M_2=200~{\rm GeV},~ \mu=-350~{\rm GeV},~M_{\ti Q}=350~{\rm GeV},\nn\\ 
   & M_{\ti Q} : M_{\ti U} : M_{\ti D} = 1:0.85:1.05,~
     A_t = A_b = -500~{\rm GeV}.
\label{eq:parset}
\end{align}
Moreover, we assume GUT relations for the gaugino mass parameters
$M_1$, $M_2$, $M_3$. In this case, the phases of the gaugino
sector can be rotated away. 
Since $\phi_\mu$, the phase of $\mu$, is highly constrained by
the EDMs of electron and neutron, we take $\phi_\mu = 0$.
%%%unless mentioned otherwise.
%
The phases relevant to our study are thus
$\phi_t$ and $\phi_b$, the phases of $A_t$ and $A_b$. 
For the choice eq.~\eq{parset}, $\tan\b=10$ and $\phi_t=0~(\pi/2)$, 
we get $\mst{1}=226$~(213)~GeV, $\mst{2}=465$~(471)~GeV, 
$\msb{1}=340$~GeV, and $\msb{2}=382$~GeV.

\begin{figure}[t!]
\setlength{\unitlength}{1mm}
\begin{center}
% --- a ---
\begin{picture}(60,62)
%\put(0,0){\framebox(60,60){}}
\put(2,0){\mbox{\epsfig{figure=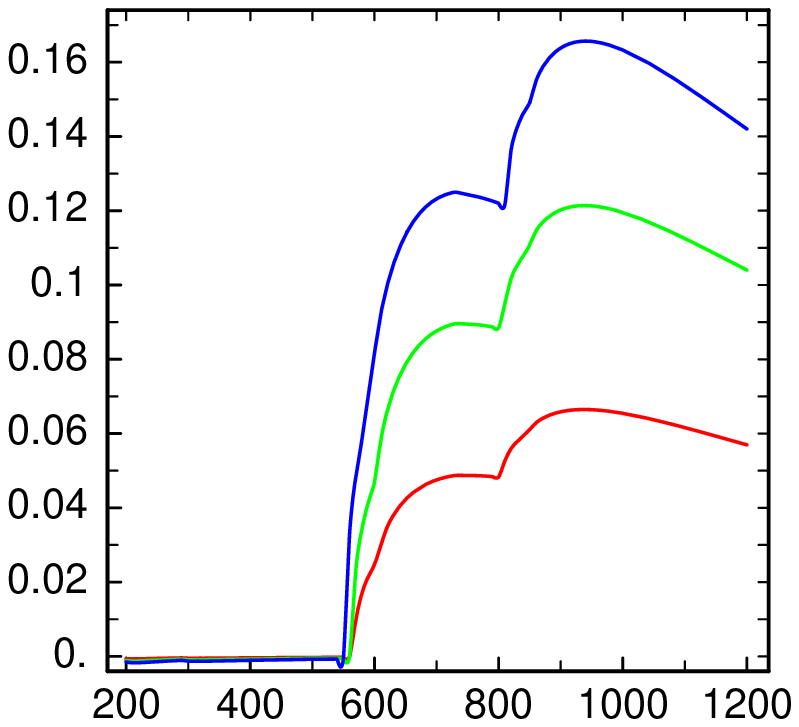,height=61mm}}}
\put(26,49){\footnotesize $\phi_t=\pi/2$}
\put(50,44){\footnotesize $\pi/4$}
\put(50,29){\footnotesize $\pi/8$}
\put(26,-2){$m_{H^+}$~[GeV]}
\put(-5,30){\rotatebox{90}{$|\d^{CP}|$}}
\put(-5,56){\bf a)}
\end{picture}
% --- b ---
\hspace{16mm}
\begin{picture}(60,62)
%\put(0,0){\framebox(60,60){}}
\put(0,0){\mbox{\epsfig{figure=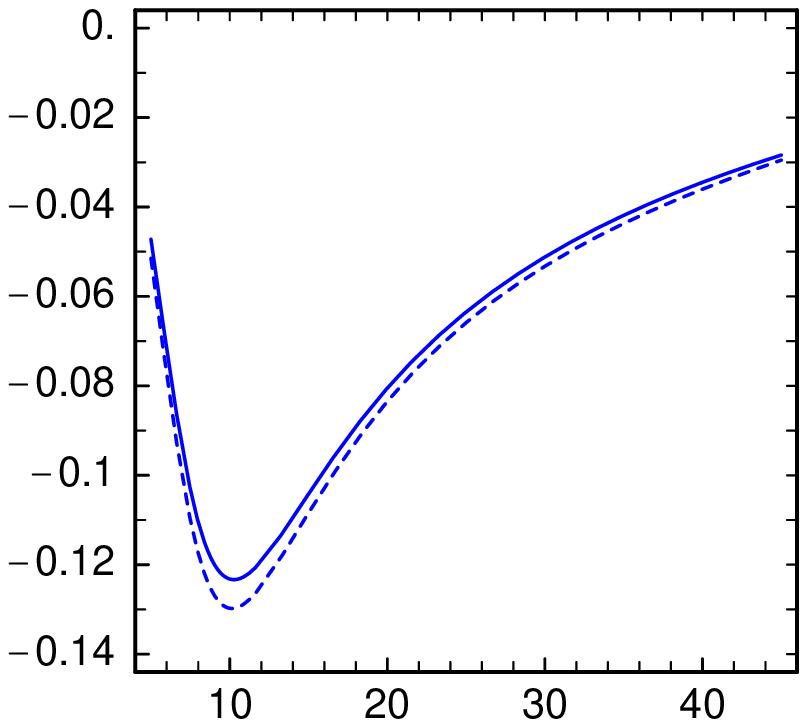,height=63mm}}}
\put(40,46){\footnotesize $\phi_b=0$}
\put(24,15){\footnotesize $\phi_b=\pi/2$}
\put(34,-2){$\tan\b$}
\put(-5,32){\rotatebox{90}{$\d^{CP}$}}
\put(-5,56){\bf b)}
\end{picture}
\caption{The decay rate asymmetry $\d^{CP}$ of $H^\pm\to tb$ 
  in {\bf (a)} as a function of $\mhp$ for $\tan\b=10$ and $\phi_b=0$, 
  in {\bf (b)} as a function of $\tan\b$, for $\mhp=700$~GeV and 
  $\phi_t=\pi/2$.
  The other parameters are fixed by eq.~\eq{parset}.
  \label{fig:S2mHtb}}
\end{center}
\end{figure}

\Fig{S2mHtb}a shows $\d^{CP}$ as a function of $\mhp$ for $\tan\b=10$.
For $\mhp < \mst{1}+\msb{1}$, $\d^{CP}$ is very small,
${\cal O}(10^{-3})$ or smaller. The contributions come from 
the diagrams of Figs.~1a, 1c, and 1f; the diagram of \fig{feyngraphs}d  
only contributes if there is a non-zero phase in the chargino/neutralino 
sector. 
However, once the $H^+\to\st\bar{\sb}$ channel is open, $\d^{CP}$ can 
go up to several per cent. The thresholds of $H^+\to\st_1\bar{\sb}_1$ at 
$\mhp\simeq 550$~GeV, and of $H^+\to\st_2\bar{\sb}_1$ at $\mhp\simeq 810$~GeV
are clearly visible in \fig{S2mHtb}a. 
For $\mhp=700$~GeV, we obtain $\d^{CP}\sim -5\%$, $-9\%$, and $-12\%$ 
for $\phi_t=\pi/8$, $\pi/4$, and $\pi/2$, respectively. 
For $\mhp=900$\,--\,1000~GeV, $\d^{CP}$ goes up to almost 17\%. 
The dominant contribution comes from the stop--sbottom--gluino loop 
of \fig{feyngraphs}b. Also the stop--sbottom--neutralino loop of Fig.~1b 
and the stop--sbottom self-energy of Fig.~1e can give a relevant contribution 
and should thus be taken into account. The contribution of the graphs with 
$\ch\nt$ or $H^0W$ (\fig{feyngraphs}a,c,f) exchange can, however, be neglected 
in this case. Here a remark is in order: 
To calculate the latter contributions with neutral Higgs bosons, we have used 
\cite{CEPW00,cph}. This is sufficient for our purpose, since we are mainly 
interested in large CP-violating effects that occur for $\mhp>\mst{1}+\msb{1}$ 
because of $\phi_{t,b}$. However, once precision measurements of $H^\pm$ decays 
become feasible, a more complete calculation of the $H^0_l$ masses and couplings 
\cite{Carena:2001fw} might be used. 

\Fig{S2mHtb}b shows the $\tan\b$ dependence of $\d^{CP}$ for $\mhp=700$~GeV 
and the cases $\phi_t=\pi/2$, $\phi_b=0$ (full line) and $\phi_t=\phi_b=\pi/2$ 
(dashed line). It turns out that the asymmetry has a maximum around 
$\tan\b\simeq 10$ and decreases %%% approaches a constant value
for larger $\tan\b$. 
For $\phi_t=\pi/2$ and $\phi_b=0$, we have $\d^{CP}\sim -12\%$ at $\tan\b=10$ 
and $\d^{CP}\sim -3.5\%$ at $\tan\b=40$. 
An additional phase of $A_b$ can enhance or reduce the asymmetry. 
%One may expect that the importance of $\phi_b$ increases with $\tan\b$. 
For the parameters eq.~\eq{parset}, however, it turns out that the 
effects in the triangle and self-energy graphs of Fig.~1b and 1e 
compensate each other so that the overall dependence on $\phi_b$ is 
small. % However, this does not hold in general.
%%%
%For completeness, we also show as a dotted line the case of a large phase 
%of $\mu$ (which is in agreement with the contraints from EDMs if there are 
%cancellations or the squarks of the first two generations are very heavy, 
%$\msq{}\gsim 10$~TeV): $\phi_\mu$ can have a large effect for small to medium 
%values of $\tan\b$, mainly through the $\st_R\sb_L H^+$ coupling 
%$\propto h_t\mu\sin\b$.
%For $\phi_t=\phi_b=\phi_\mu=\pi/2$ and $\tan\b=10$, $\d^{CP}\sim -20\%$. 

% ---------- Figure 2 ----------

% ---------- Figure 3 ----------
\begin{figure}[t!]
\setlength{\unitlength}{1mm}
\begin{center}
% --- a ---
\begin{picture}(60,62)
%\put(0,0){\framebox(60,60){}}
\put(0,0){\mbox{\epsfig{figure=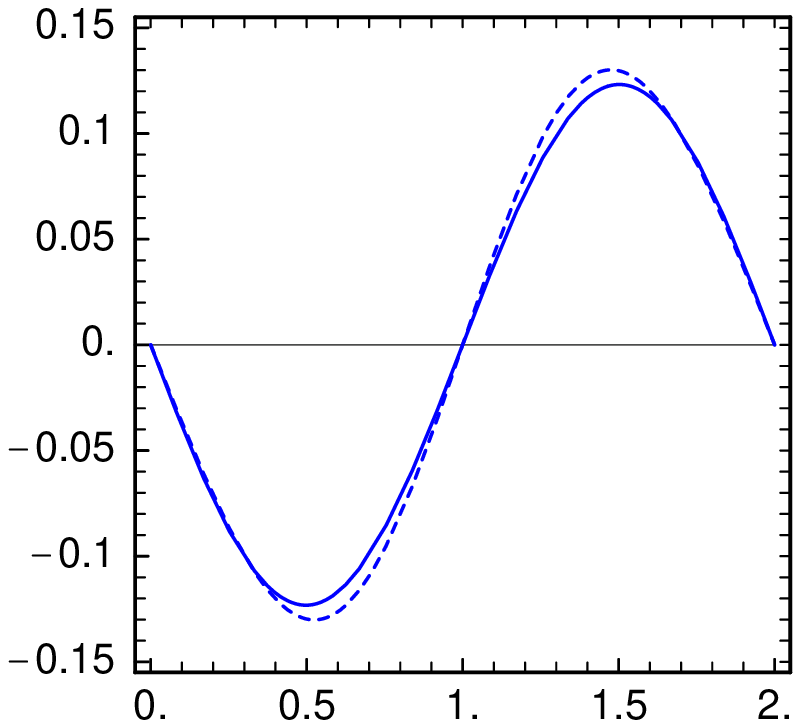,height=63mm}}}
\put(40,13){\footnotesize\mbf $\tan\b=10$}
\put(32,-2){$\phi_t~[\pi]$}
\put(-5,31){\rotatebox{90}{$\d^{CP}$}}
\put(-5,56){\bf a)}
\end{picture}
% --- b ---
\hspace{16mm}
\begin{picture}(60,62)
%\put(0,0){\framebox(60,60){}}
\put(0,0){\mbox{\epsfig{figure=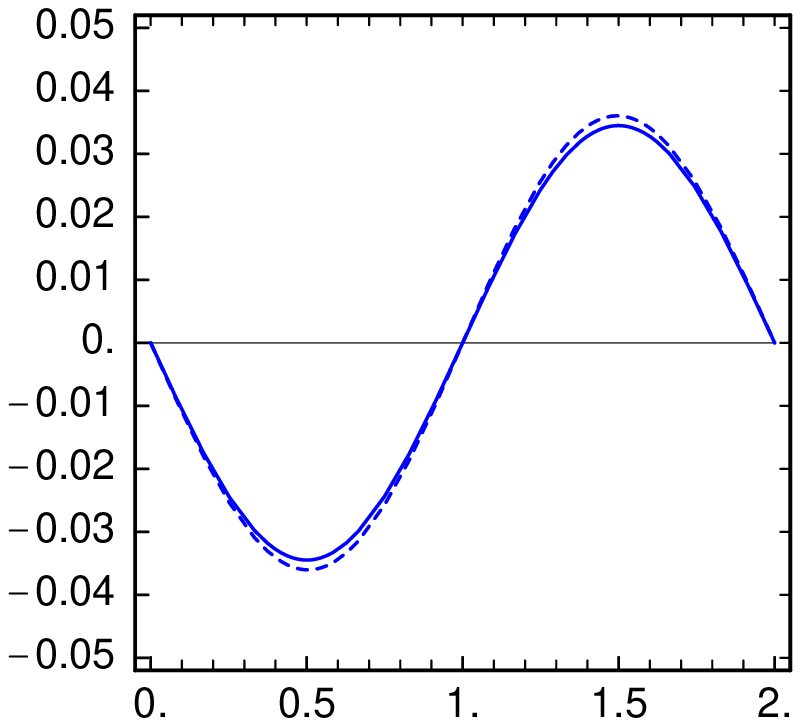,height=63mm}}}
\put(40,13){\footnotesize\mbf $\tan\b=40$}
\put(32,-2){$\phi_t~[\pi]$}
\put(-5,31){\rotatebox{90}{$\d^{CP}$}}
\put(-5,56){\bf b)}
\end{picture}
\caption{$\d^{CP}$ as a function of $\phi_t$, for $\mhp=700$~GeV, 
  $\tan\b=10$ in {\bf (a)} and $\tan\b=40$ in {\bf (b)};
  full lines: $\phi_b=0$, dashed lines: $\phi_b=\phi_t$.
  The other parameters are fixed by eq.~\eq{parset}.
  \label{fig:S2At}}
\end{center}
\end{figure}

The dependence on $\phi_t$ is shown explicitly in \fig{S2At}, where we plot 
$\d^{CP}$ as a function of $\phi_t$, for $\mhp=700$~GeV and $\tan\b=10$ 
and 40.
As expected, $\d^{CP}$ shows a $\sin\phi_t$ dependence.
%Note again the influence of $\phi_b$: in case of a maximal stop phase,
%$\phi_b$ can change $\d^{CP}$ by up to 20\% for $\tan\b=10$, and
%by up to 40\% for $\tan\b=40$.
Here note that the branching ratio of $H^+\to t\bar b$ increases 
with $\tan\b$. For $\mhp=700$, in the case of vanishing phases, we have 
BR$(H^+\to t\bar b)\simeq 17\%$ (85\%) for $\tan\b=10$ (40).

\begin{figure}[h!]
\setlength{\unitlength}{1mm}
\begin{center}
\begin{picture}(60,64)
%\put(0,0){\framebox(60,60){}}
\put(0,0){\mbox{\epsfig{figure=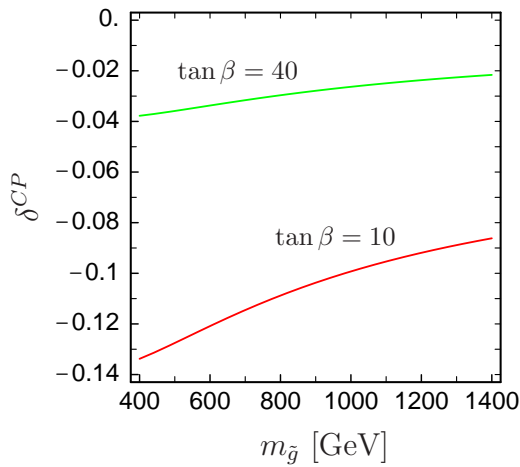,height=63mm}}}
\put(30,27){\footnotesize $\tan\b=10$}
\put(17,49){\footnotesize $\tan\b=40$}
\put(28,-1){$\msg$~[GeV]}
\put(-5,30){\rotatebox{90}{$\d^{CP}$}}
\end{picture}
\caption{$\d^{CP}$ as a function of $\msg$, for $\mhp=700$~GeV, 
  $\phi_t=\pi/2$, $\phi_b=0$, $\tan\b=10$ and 40. 
  The other parameters are fixed by eq.~\eq{parset}.
  \label{fig:S2msg}}
\end{center}
\end{figure}

Last but not least we relax the GUT relations between the gaugino 
masses and take $\msg$ as a free parameter (keeping, however, the 
relation between $M_1$ and $M_2$ and taking $M_3=\msg$ real). 
\Fig{S2msg} shows the dependence of $\d^{CP}$ on the gluino mass
for $\mhp=700$~GeV, $\phi_t=\pi/2$, $\phi_b=0$, and $\tan\b=10$ and 40. 
As one can see, the gluino does not decouple. Even for a large 
gluino mass, an asymmetry of a few per cent is possible. 
A non-zero phase of $M_3$ may also have a large effect.  
It can, in fact, lead to an asymmetry of ${\cal O}(10\%)$ 
even if all other phases are zero \cite{Christova:2002ke}.

% ---------- Figure 4 ----------

Some remarks on the measurability are in order. 
At the Tevatron, no sensitivity for detecting $H^\pm$ is 
expected for a mass $\mhp\gsim 200$~GeV. 
The LHC, on the other hand, has a discovery reach up to $\mhp\sim 1$~TeV, 
especially if QCD and SUSY effects conspire to enhance the cross 
section. With a luminosity of ${\cal L}=100$~fb$^{-1}$, 
about 217 signal events can be expected for $pp\to H^+\bar t b$ with 
$S/\sqrt{B}=6.3$, for $\mhp\simeq 700$~GeV and 
$\tan\b=50$~\cite{Belyaev:2002eq}. However, the 
region $\tan\b\lsim 20$ seems to be very difficult.  
In $e^+e^-$ collisions, the dominant production mode is $e^+e^-\to H^+H^-$. 
For the mass ranges relevant to our study this would require a TeV-scale 
linear collider. Indeed, for $\mhp\sim 700$~GeV, BR$(H^\pm\to tb)$ could 
be measured to few per cent at CLIC \cite{Battaglia:2001be}.

To summarize, we have calculated the difference between the partial 
decay rates $\G\,(H^+ \to t\bar b)$ and $\G\,(H^- \to \bar tb)$ due to 
CP-violating phases in the MSSM. 
The resulting rate asymmetry $\d^{CP}$, eq.~(1), could be measured in a 
counting experiment. If $\mhp < \mst{1}+\msb{1}$, $\d^{CP}$ is typically 
of the order of $10^{-3}$. However, for $\mhp > \mst{1}+\msb{1}$, $\d^{CP}$ 
can go up to 10\,--\,15\%, depending on the phases of $A_t$, $A_b$, 
and $\mu$, and on $\tan\b$. Such a large asymmetry should be measurable 
at future colliders such as LHC or CLIC.

%==============================================================================
%  Bibliograpy
%==============================================================================

\end{document}